\documentstyle[12pt,epsf,psfig]{article}

\large
\newcommand{\be}{\begin{eqnarray}}
\newcommand{\ee}{\end{eqnarray}}
\newcommand\del{\partial}

\begin{document}
\setlength{\baselineskip}{21pt}
\pagestyle{empty}
\vfill
\eject
\begin{flushright}
SUNY-NTG-95/26
\end{flushright}

\vskip 2.0cm
\centerline{\Large A random matrix model for chiral symmetry breaking}
\vskip 2.0 cm
\centerline{\bf A.D. Jackson and J.J.M. Verbaarschot}
\vskip .2cm
\centerline{Department of Physics}
\centerline{SUNY, Stony Brook, New York 11794}
\vskip 2cm

\centerline{\bf Abstract}
We formulate a random matrix model which mimics the chiral phase
transition in QCD with two light flavors.
Two critical exponents are calculated. We obtain
the mean field values $\beta = \frac 12 $ and $\delta = 3$. We also
find that the chiral phase transition can be characterized by the
dynamics of the smallest eigenvalue of the Dirac operator. This suggests an
alternative order parameter which may be of relevance for lattice QCD
simulations.

\vfill
\noindent
\begin{flushleft}
SUNY-NTG-95/26\\
August 1995
\end{flushleft}
\eject
\pagestyle{plain}

\vskip 1.5cm
\noindent
\renewcommand{\theequation}{1.\arabic{equation}}
\setcounter{equation}{0}
\section{Introduction}
\vskip 0.5cm
In recent years the QCD phase transition has been studied in a variety of
ways using both numerical simulations and analytical methods.  Such studies
have led to the conviction that it may be a second-order phase transition
\cite{Karsch,Christ}. As has been stressed, in particular in
\cite{Pisarski-Wilczek,Rajagopal-Wilczek}, this has important consequences
because the transition can then be characterized
by critical exponents corresponding to a specific universality class.
In particular it was argued that the critical exponents are those of
an $O(4)$ Heisenberg spin model.  (On a lattice and with Kogut-Susskind
fermions, the relevant model might rather be an $O(2)$ spin model.)
However, as always, universality arguments  must be used with care.
According to a recent suggestion in \cite{Kocic-Kogut} based on simulations
of the three dimensional Gross-Neveu model, they  may not be valid
for phase transitions involving soft modes composed of fermions.
The reason is that the lowest Matsubara frequency suppresses
infrared divergencies  which lead to the universal critical exponents.
Indeed, Kocic and Kogut found that the critical exponents in their
model are given by mean field theory.

We wish to study the chiral phase transition from the perspective of the
spectrum of the Euclidean Dirac operator.   Although initial numerical lattice
results regarding this issue have  become
available \cite{Hands-Teper,Kalkreuter,Christ}, a systematic study is beyond
the
reach of present day computers.   Thus, we prefer to construct a simple
model which contains the main ingredients of QCD related to chiral symmetry
breaking.  The zero temperature version of this model was considered
previously \cite{V}, and it was shown \cite{Halasz-V} that it is equivalent
to the finite volume effective partition function. In particular, it was
shown \cite{V} that the spectrum of the Dirac operator in this random matrix
model obeys the so-called Leutwyler-Smilga sum rules \cite{LS}.

In this work we consider the non-zero temperature results of two versions of
the random matrix model, for $SU(2)$ and for $SU(N_c), \,\, N_c \ge 3$ both
with fundamental fermions, which are characterized by real and complex matrix
elements, respectively. We shall give analytical results for the critical
exponents $\beta$ and $\delta$ as well as numerical results (for the case
of complex matrices only).  We also study the temperature dependence of
the spectrum.  Finally, we calculate the dependence of the chiral condensate
on the valence quark mass. These results will be shown to be in qualitative
agreement with recent lattice calculations \cite{Christ}.

\vskip 1.5 cm
\renewcommand{\theequation}{2.\arabic{equation}}
\setcounter{equation}{0}
\section{Finite temperature chiral random matrix model}
\vskip 0.5 cm

The QCD partition function for vacuum angle $\theta$ is defined as
\be
Z_{QCD} = \sum_\nu e^{i\nu\theta} Z_{QCD}(\nu)\ \ ,
\ee
where the partition function in a sector with topological charge $\nu$ and
$N_f$ fermionic flavors is given
by
\be
Z_{QCD}(\nu)= \langle \prod_{f=1}^{N_f} m_f^\nu \prod_{\lambda_n > 0}
(\lambda_n^2 +m^2_f)\rangle_A \ \ .
\ee
Here, $\langle\cdots\rangle_A$ denotes the averaging over the gauge
field configurations, with topological charge $\nu$,
weighted according the the QCD action.
The eigenvalues, $\lambda_k$, of the Dirac operator fluctuate over the ensemble
of gauge field configurations. In general, the complete eigenvalue density
is a determined in a nontrivial way by the QCD dynamics. However, fluctuations
of the eigenvalues on the microscopic scale (on the scale of the average
level spacing) are believed to show universal characteristics. It is our
conjecture that the eigenvalues near zero virtuality obey such microscopic
universality. This implies that the detailed dynamics of the QCD partition
function are not important for such fluctuations, and can be described
equally well with a random matrix ensemble which respects
the global symmetries of the Dirac operator.
In particular, the following properties are included \cite{V}:
(i) The $U_A(1)$
symmetry leading to an eigenvalue spectrum $\pm \lambda_n$.   (ii)
The zero mode structure of the Dirac operator.  (In the sector of
topological charge $\nu$ the Dirac operator has exactly $\nu$ zero
eigenvalues all of the same  chirality.)  (iii) The flavor
chiral symmetry and its spontaneous
or explicit  breaking.  (iv) The reality type of the representation of the
gauge group.  For $SU(N_c), \,\,N_c\ge 3$ in the fundamental
representation, the gauge field is complex and
so are the matrix elements of the Dirac operator. The gauge group $SU(2)$
in the fundamental representation is pseudoreal leading to matrix elements
of the Dirac operator that are real. Finally, for gauge group $SU(N_c)$ in
the adjoint representation, the gauge field is real and the matrix elements
of the Dirac operator can be organized into real quaternions.

In this work we want to construct a model which describes the
fluctuations of the smallest eigenvalues as a function of the temperature.
Near $T_c$ the theory of critical phenomena tells us that the fluctuations
are universal with non-trivial critical exponents  which are related to
the propagation of soft modes. However,
 the recent work of Kocic and Kogut \cite{Kocic-Kogut}
suggests that this scenario may not be valid for
phase transitions involving bosons composed of fermions.  Instead,
the lowest Matsubara frequency, $\pi T$, cures the infrared divergences
and leads to a mean-field-like second-order phase transition.
According to their work on the  three dimensional Gross-Neveu model, the
dynamics of the phase transition are determined by the
lowest Matsubara frequency.  In this spirit, the only temperature effect we
include in our model is
that of the lowest Matsubara frequency between each pair of
suitably chosen basis states $(1\pm\gamma_5)\, \phi_n$. In the sector
of topological charge $\nu$ our basis  must be complemented
by $\nu$ unpaired basis states of the same chirality. Together with the
symmetries mentioned above, this leads to the random matrix model,
\be
Z_\beta(\nu, N_f) = \int {\cal D} W {\det}^{N_f} \left (
\begin{array}{cc} {\bf m}^* & iW+i\pi T \\ iW^\dagger +i\pi T & {\bf m}
\end{array} \right )
\exp[-\frac{n\beta\Sigma^2}2 \,{\rm Tr}\, W W^\dagger] \ \ ,\nonumber\\
\label{zrandom}
\ee
where $W$ is an $n\times (n+\nu)$ matrix. The integration over $W$ is to
be performed according to the Haar measure.  We also include an
arbitrary complex mass matrix ${\bf m}$ with mass eigenvalues equal to $m_f$.
For QCD with $N_c = 2$ the matrix elements of $W$ are real
($\beta =1$).  They are complex for $N_c \ge 3$ ($\beta = 2$).  In each
case we include fermions in the fundamental representation.

The fermion determinant in (\ref{zrandom})
can be written as a Grassmann integral.
\be
Z_\beta(\nu, N_f) =
\int {\cal D} W {\cal D} \psi^* {\cal D} \psi
\exp[i\sum_{k=1}^{N_f}\psi^{k\,*} \left (
\begin{array}{cc} -i{\bf m}^* & W + \pi T \\ W^\dagger +\pi T& -i{\bf m}
\end{array} \right )
\psi^k   \exp[-\frac{n\beta\Sigma^2}2 \,{\rm Tr}\, WW^\dagger] \ \ .\nonumber
\\
\label{ranpart}
\ee
The quenched approximation is obtained from this model as in the replica
trick. We calculate a property for arbitrary $N_f$ and take the limit
$N_f\rightarrow 0$ at the end of the calculation. In particular, quantities
which are $N_f$-independent are valid for $N_f = 0$ as well.

It should be stressed that the partition function (\ref{zrandom}) represents
a {\em schematic\/} model for the chiral phase transition.  Although the
temperature dependence of this model does not {\em coincide\/} with that of
chiral perturbation theory \cite{Gasser-Leutwyler}, it will be shown below
that there is considerable {\em qualitative\/} agreement.

\vskip 1.5 cm
\renewcommand{\theequation}{3.\arabic{equation}}
\setcounter{equation}{0}
\section{Analytical results}
\vskip 0.5 cm
In this section we evaluate the partition function (\ref{ranpart}) using
methods which are standard in the supersymmetric formulation of random
matrix theory \cite{Efetov,VWZ}.  The first step is to perform
the average over $W$ by performing a Gaussian integral. This leads
to a four-fermion interaction.  After averaging over the matrix
elements of the Dirac operator, the partition function becomes
\be
Z_1(\nu, N_f) =
\int {\cal D} \psi^* {\cal D} \psi\exp[&-&\frac 1{2n\Sigma^2 \beta}
(\psi^{f\,*}_{R\, i}\psi^{f}_{L\, k}
\psi^{g\,*}_{R\, i}\psi^{g}_{L\,k }
+2\psi^{f\,*}_{R\, i}\psi^{f}_{L\, k}
\psi^{g\,*}_{L\, k}\psi^{g}_{R\, i }
+\psi^{f\,*}_{L\, k}\psi^{f}_{R\, i}
\psi^{g\,*}_{L\, k}\psi^{g}_{R\,i }) \nonumber\\
&+ & {\bf m}_{fg}^*\psi^{f\,*}_{R\, i}
\psi^{g}_{R\, i}+{\bf m}_{fg}\psi^{f\,*}_{L\, k}\psi^{g}_{L\, k}
+i\pi T(\psi^{f\,*}_{R\, i} \psi^{f}_{L\, i}
+\psi^{f\,*}_{L\,k}\psi^{f}_{R\, k})] \ \ ,
\ee
for $\beta = 1$, and
\be
Z_2(\nu, N_f) = \int {\cal D} \psi^* {\cal D}\psi \exp[ -\frac
2{n\Sigma^2 \beta} \psi^{f\,*}_{L\, k}\psi^{f}_{R\, i} \psi^{g\,*}_{R\,
i}\psi^{g}_{L\,k }&+&{\bf m}_{fg}^*\psi^{f\,*}_{R\, i} \psi^{g}_{R\, i}
+{\bf m}_{fg}\psi^{f\,*}_{L\,
k}\psi^{g}_{L\, k}\nonumber \\
 &+&i\pi T(\psi^{f\,*}_{R\, i} \psi^{f}_{L\, i}
+\psi^{f\,*}_{L\,k}\psi^{f}_{R\, k})]
\ee
for $\beta = 2$.  In both cases, each of the four-fermion terms can be
written as the difference of two squares. Each square can be linearized by
the Hubbard-Stratonovitch transformation according to
\be
\exp(-AQ^2) \sim \int d\sigma\exp(-\frac{\sigma^2}{4A} - iQ \sigma) \ \ .
\label{Hubbard}
\ee
For $\beta = 2$ the partition function, expressed in terms of the two bosonic
variables, can be combined into a single complex $N_f\times N_f$ matrix, $A$,
resulting in
\be
Z_2(\nu, N_f) &=&
\int {\cal D} A
{\cal D} \psi {\cal D} \psi^* \exp[ -\frac{n\Sigma^2\beta}2 {\rm Tr}
A A^\dagger \nonumber\\ &+&\psi^{f\,*}_{L\,k}
\psi^{g}_{L\,k}(A +{\bf m}) +\psi^{f\,*}_{R\,i} \psi^{g}_{R\,i}(A^\dagger
+{\bf m}^*)
+i\pi T(\psi^{f\,*}_{R\, i} \psi^{f}_{L\, i}
+\psi^{f\,*}_{L\,k}\psi^{f}_{R\, k})] \ \ .
\ee
For $\beta=1$ six new bosonic matrix variables are required.  This is
related to the fact that, for two colors, baryons are composed of two quarks
and are bosons. They can be combined into one antisymmetric, complex
$2 N_f \times 2 N_f$ matrix $A$ resulting in the partition function
\be
&&Z_1(\nu, N_f) =
\int {\cal D} A{\cal D} \psi {\cal D} \psi^* \exp[ -\frac{n\Sigma^2 \beta}2
{\rm Tr} A A^\dagger] \nonumber\\
&&\times\exp \frac 12 \left (\begin{array}{c} \psi_R \\ \psi_R^*
\end{array} \right ) (A^\dagger + {\cal M}^*)
\left (\begin{array}{c}\psi_R \\ \psi_R^* \end{array}\right) \nonumber \\
&&\times\exp \frac 12\left
(\begin{array}{c}\psi_L\\ \psi_L^* \end{array}\right)
\left ( \begin{array}{cc} 0 & -{\bf 1}\\ {\bf 1} & 0 \end{array}
\right ) (-A +\cal {M})
\left ( \begin{array}{cc} 0 & {\bf 1}\\ -{\bf 1} & 0 \end{array} \right )
\left (\begin{array}{c} \psi_L \\ \psi_L^* \end{array}\right)\nonumber \\
&&\times\exp[i\pi T(\psi^{f\,*}_{R\, i} \psi^{f}_{L\, i}
+\psi^{f\,*}_{L\,k}\psi^{f}_{R\, k})].
\label{realall}
\ee
In this case the mass matrix is an antisymmetric matrix given by
\be
{\cal M} = \left ( \begin{array}{cc} 0 & -{\bf m}\\ {\bf m} & 0 \end{array}
\right ) \ \ .
\ee
Note also that the temperature-dependent term can be rewritten as
\be
\frac{i\pi T}2\left (\begin{array}{c}\psi_L\\ \psi_L^* \end{array}\right)
\left ( \begin{array}{cc} 0 & -{\bf 1}\\ {\bf 1} & 0 \end{array}
\right )
\left (\begin{array}{c} \psi_R \\ \psi_R^* \end{array}\right)
- (L \longleftrightarrow R)
\ee

Using this, the fermionic integrals can be performed, and the
partition function is given by
\be
Z_2(\nu, N_f) = \int {\cal D} A \exp [-\frac{n\Sigma^2\beta}2 {\rm Tr}
A A^\dagger] {\det}^{|\nu|} (A+{\bf m} ){\det}^{n}
\left ( \begin{array}{cc}  A+{\bf m} & \pi i T \\  \pi i T &A^\dagger +{\bf
m^*}
\end{array} \right )
\label{apart2}
\ee
for $\beta = 2$, and
\be
Z_1(\nu, N_f) = \int {\cal D} A \exp [-\frac{n\Sigma^2\beta}2 {\rm Tr}
A A^\dagger] {\rm Pf}^{|\nu|} (-A+{\cal M} ){\rm Pf}^{n}
\left ( \begin{array}{cc}  A^\dagger+{\cal M}^* & \pi i T \\
-\pi i T & -A +{\cal M}
\end{array} \right )
\label{apart1}
\ee
for $\beta =1$. In (\ref{apart2}) $A$ is an arbitrary complex matrix whereas
in (\ref{apart1}) $A$ is an arbitrary antisymmetric complex matrix.

In each case the condensate is given by
\be
\langle \bar q q \rangle = \frac 1{2n N_f} \del_{m} \log Z \ \ ,
\ee
where $Z$ is evaluated for a diagonal mass matrix with equal diagonal
matrix elements.  In the limit $n \rightarrow \infty$,  the
condensate can be determined with the aid
of a saddle point approximation. The saddle point equations for
$\beta = 2$ are given by
\be
-\frac{n\beta \Sigma^2}2 A+ n(A+m)\left( (A^\dagger +m)(A+m) + \pi^2 T^2\right
)^{-1} = 0 \ \ \ .
\label{spe}
\ee
An arbitrary complex matrix can be diagonalized by performing the
decomposition
\be
A= U \Lambda V^{-1},
\ee
with all eigenvalues positive and $U$ and $V$ unitary matrices.
We find that the solution of (\ref{spe}) yields $U = V = 1$ with
eigenvalues $\lambda$ given by the positive roots of
\be
\Sigma^2\lambda ((\lambda+m)^2 + \pi^2 T^2) - \lambda - m = 0  \ \ .
\ee
In the chiral limit we find a critical point at
\be
T_c = \frac 1{\pi \Sigma} \ \ \ .
\ee

In order to calculate the condensate, we express the derivative of the
partition function in (3.10) in terms of an average over $A$,
\be
\langle \bar q q \rangle =  \frac 1{2n N_f} \langle{\rm Tr} \left (
\begin{array}{cc} A^\dagger & \pi iT \\ \pi i T & A \end{array}\right )^{-1}
\rangle \ \ .
\ee
Below $T_c$ the mass $m$ can be neglected in the saddle point
equation, and we find
\be
\langle \bar q q \rangle = \Sigma (1-\pi^2T^2\Sigma^2)^{1/2} \ \ .
\label{cond1}
\ee
At $T_c$ the solution of the saddle point equation develops a non-analytic
dependence on $m$ resulting in the condensate
\be
\langle \bar q q \rangle = \Sigma^{\frac{4}{3}} m^{\frac{1}{3}} \ \ .
\ee
Therefore, we reproduce the mean field value for the critical exponent
$\delta = 3$.

For $\beta =1$ the saddle point equation for the $A$ integration is
\be
-\frac{\beta\Sigma^2}2 A + \frac 12(A+{\cal M})\left( (A^\dagger +{\cal M})
(A-{\cal M}) + \pi^2 T^2\right
)^{-1} = 0 \ \ .
\label{spe1}
\ee
This equation can be solved by diagonalizing the complex
antisymmetric matrix
$A$ as $A = U \Lambda \tilde U$, where $U$ is a unitary matrix.  Here,
$\Lambda$ is an antisymmetric, standard matrix such that $\Lambda_{k, k+1}
= - \Lambda_{k+1, k} = \lambda_k$ for $k = 1, \, \cdots, 2N_f -1$ with all
other matrix elements equal to zero.  A suitable re-definition of $U$ can
always made such that all $\lambda_k \ge 0$.  The condensate is calculated
as in the case of $\beta =2 $ with the same result both below and at $T_c$.
Of course, it comes as no surprise that we obtain the mean field
value for the critical exponent in this case as well.

\vskip 1.5 cm
\renewcommand{\theequation}{4.\arabic{equation}}
\setcounter{equation}{0}
\section{The phase transition}

In the remainder of this paper we describe the results of numerical
investigations of the random matrix model (2.3).  For $N_f\ne 0$, the
determinantal weight must be included
in the integration measure which is extremely costly. However, for $N_f=0$
the distribution functions are simple Gaussians, and
only this case will be
studied.  Instead of a Gaussian distribution of matrix elements, we
have used a rectangular distribution centered about zero with variance given
by
\be
\overline{|W_{ij}|^2} = \frac 1{6n} \ \ ,
\ee
where $n$ is the size of the off-diagonal blocks. If we compare this to the
variance of the Gaussian distribution (2.3), we
can make the identification
\be
\frac 1{6n} = \frac 2{n\beta \Sigma^2},
\ee
which yields $\Sigma^2 = 6$ for $\beta = 2$.
According to (3.14) the critical
temperature is thus $\pi T_c = 1/\sqrt 6 = 0.40824$.   General universality
arguments \cite{univers} imply that the shape of the distribution of the
matrix elements does not affect the properties of our random matrix model,
and we expect that the results for the rectangular distribution will be
in complete agreement with those for the Gaussian distribution.

The chiral order parameter of our schematic model is the spectral density
$\rho(0)$. It is related the the chiral condensate {\em via\/} the Banks-Casher
formula \cite{BANKS-CASHER}
\be
\langle \bar q q \rangle = \frac{1}{2n} \pi \rho(0) \ \ .
\ee
Let us first consider the complete spectral density $\rho(\lambda)$.
At $T = 0$ it can be shown, using arguments
familiar from random matrix theory \cite{MEHTA}, that this density
has a semicircular shape. For non-zero temperature, it is somewhat
more difficult to obtain the level density analytically \cite{JSV}.
In Figs.\ 1a, 1b, and 1c, we show numerical results for the complete spectral
density for $\pi T= 0$, $\pi T= 0.4$, and $\pi T = 1.0$.  The deviations
from a semi-circle at $T=0$ are not statistical fluctuations but rather
well-understood finite-$n$ corrections \cite{VZh}.
When $T$ is large,
we find that the distribution splits into two semicircles with centers
at $+\pi T$ and $-\pi T$.
It is elementary to demonstrate that all eigenvalues come
in pairs located symmetrically about zero.  When $T=0$, the distribution
of the eigenvalue of smallest magnitude is also known analytically.
For the moment, we note only that $ \rho(\lambda) \sim \lambda$ for values
of $\lambda$ less than the average value of this eigenvalue of smallest
magnitude.  Thus, some care is required in order to extract a meaningful
({\em i.e.}, non-zero) value of $\rho(0)$.  The simplest method is to construct
$\rho( 0 )$ from the fraction of eigenvalues in an interval $2 \, \Delta
\lambda$ centered about $0$.  When $\Delta \lambda$ is less than the average of
eigenvalue of smallest magnitude, this estimate of $\rho (0)$ is proportional
to $\Delta \lambda$.
\vskip 0.5 cm
\noindent
\begin{figure}[h]
\begin{center}
\leavevmode
\psfig{file=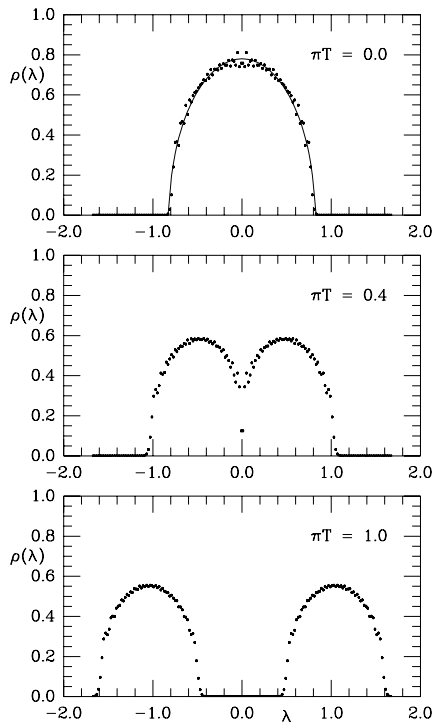,height=9.5cm}
\end{center}
\oddsidemargin3.7cm
\evensidemargin3.7cm
\setlength{\textwidth}{16cm}
\caption{Histograms of the complete spectrum of eigenvalues obtained
for $n=20$ at temperatures of (a) $\pi T = 0$ (top curve), (b) $\pi T = 0.4$
(middle curve), and (c) $\pi T = 1.0$ (bottom curve).
Each spectrum was obtained from $10^5$ matrices.}
\oddsidemargin1.7cm
\evensidemargin1.7cm
\setlength{\textwidth}{18cm}
\end{figure}

\noindent
For somewhat larger values of $\Delta \lambda$ there is
a relatively rapid saturation of $\rho ( 0 )$ to its desired asymptotic value.
In principle,it is better to obtain $\rho ( 0 )$
by performing a smearing of the
full spectrum with a Gaussian whose width is roughly comparable to the
spacing between adjacent eigenvalues.  This avoids the small-amplitude
oscillations associated with the first approach.  In practice, there is little
difference between the two methods.

According to the mean field argument presented in section 3, our model shows
a second-order phase transition with
\be
\rho(0,T) \sim \sqrt{T_c^2 -T^2} \ \ .
\ee
In order to account for finite $n$ effects approximately, we convolute
this expression with a Gaussian
\be
\rho_\sigma(0,T) \sim \int_{-T_c}^{T_c} dx \sqrt{T_c^2 - x^2}
\exp[- (x-T)^2/\sigma^2] \ \ .
\ee
While we offer no analytic justification for this form, we note that it
provides an excellent fit to the results of simulations.
A best fit of this expression to our results for $n = 20$ is shown in
Fig.\ 2.  The spectral density was obtained from a bin size of $0.02$.
In this case, the fitted values for $T_c$ and $\sigma$ are $0.389$ and
$0.074$, respectively.  Using data for $n \ge 50$ we performed
an extrapolation to $n\rightarrow
\infty$.  We find a critical temperature of $T_c = 0.3967$ which is in good
agrees with the theoretical value of $\pi T_c= 1/\sqrt 6$ for our present
numerical parameters.

\vskip 0.5 cm
\noindent
\begin{figure}[h]
\begin{center}
\leavevmode
\psfig{file=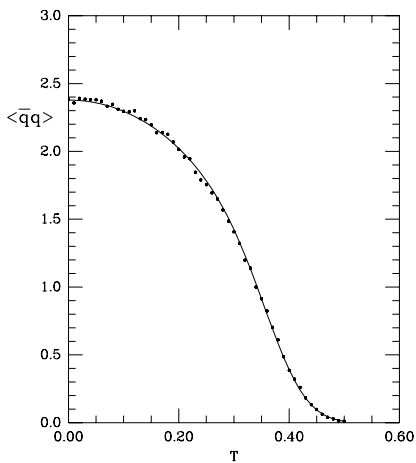,height=6cm}
\end{center}
\oddsidemargin3.7cm
\evensidemargin3.7cm
 \setlength{\textwidth}{16cm}
\caption{  The $\langle {\bar q}q \rangle$ condensate as calculated from
(4.3) for $n = 20$.  Each point was obtained on the basis of $2 \times 10^4$
matrices, and $\rho ( 0 )$ was obtained by counting eigenvalues in the range
$0 \le \lambda \le 0.02$.  The solid curve is a fit according to (4.5)
using the parameters $T_c = 0.3890$ and $\sigma = 0.7414$.  }
\oddsidemargin1.7cm
\evensidemargin1.7cm
\setlength{\textwidth}{18cm}
\end{figure}

Recent lattice QCD calculations \cite{Christ} have studied the condensate
as a function of the so-called valence mass.  In these lattice simulations, the
sea-quark mass was much larger than the smallest eigenvalue so that
the problem is effectively equivalent to taking $N_f = 0$.  It
is possible to simulate these
calculations within the framework of the present model by defining
\be
\langle \bar \psi \psi \rangle = \frac 1n \sum_{k=1}^n \frac {m}{\lambda_k^2
+m^2} \ \ .
\ee

\vskip 0.5 cm
\noindent
\begin{figure}[h]
\begin{center}
\leavevmode
\psfig{file=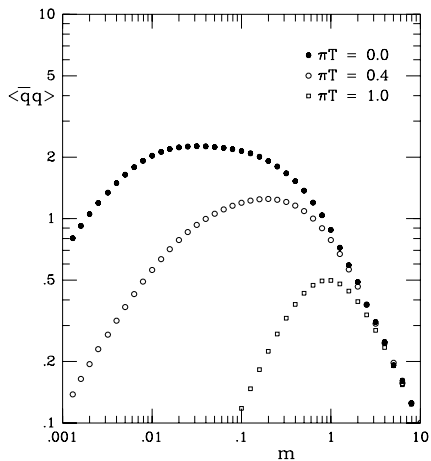,height=6cm}
\end{center}
\oddsidemargin3.7cm
\evensidemargin3.7cm
 \setlength{\textwidth}{16cm}
\caption{  The condensate as a function of the so-called valence
quark mass calculated according to (4.6) at $\pi T = 0$, $\pi T = 0.4$
and $\pi T = 1.0$.  The calculations were performed for $n=20$ and
$2 \times 10^4$ matrices were used for each temperature.  }
\oddsidemargin1.7cm
\evensidemargin1.7cm
\setlength{\textwidth}{18cm}
\end{figure}
\noindent
Fig.\ 3 shows a log-log plot of the
resulting values for $\langle \bar \psi \psi \rangle$
as a function of $m$ for $n = 20$.  Our results are qualitatively similar to
the lattice calculations.  The $m \rightarrow 0$ limit is clearly ambiguous.
These results show that the spectral density yields a much more
accurate determination of the chiral condensate \cite{V1}.

\vskip 1.5 cm
\renewcommand{\theequation}{5.\arabic{equation}}
\setcounter{equation}{0}
\section{Dynamics of the smallest eigenvalue}
\vskip 0.5 cm
In this section we study the distribution of the eigenvalue of smallest
magnitude, denoted by $\lambda_1$,
 as a function of temperature. This distribution is known
analytically for a few special cases of the Laguerre ensemble.
The present
case, defined by the random matrix model (2.3) for $N_f =0$, at $T = 0$
happens to be one of them.  With our choice of parameters, the distribution
is given by \cite{Forrester}
\be
f(\lambda_1) \sim \lambda_1 \exp{\left( - \left( \frac{n\beta \, \Sigma
\, \lambda_1 }{2}2 \right)^2 \right)} \ \ .
\ee
This $T = 0$ result is exact
even for finite $n$.  (The factor of $\lambda_1$ is readily understood.
For every eigenvalue $+ \lambda$ there is always a corresponding eigenvalue of
$- \lambda$.  This factor is simply a consequence of level repulsion.)
In order to be able to describe the temperature dependence, we introduce
a more general distribution
\be
f(\lambda_1) = \lambda_1\ \exp{\left(
-\frac{(\lambda_1 - x_0)^2}{\sigma^2} \right)} \ \ .
\ee
At zero temperature $x_0 = 0$ and $\sigma = 2/n \, \beta \, \Sigma$. The
average value of the smallest eigenvalue is given as
$\overline{\lambda_1} = \sqrt{\pi}/n \, \beta \, \Sigma$.  While this
expression is not, in general, exact, it is serviceable.

\vskip 0.5 cm
\noindent
\begin{figure}[h]
\begin{center}
\leavevmode
\psfig{file=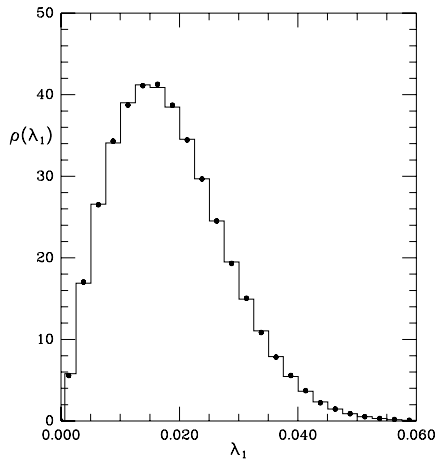,height=6cm}
\end{center}
\oddsidemargin3.7cm
\evensidemargin3.7cm
 \setlength{\textwidth}{16cm}
\caption{   The distribution of the eigenvalue of smallest magnitude
determined from $2 \times 10^5$ trials with $n=20$ for a bin size
of $0.02$ (points).  The histogram was obtained from (5.2) with $x_0 = 0$.
The value
of $\sigma = 0.020685$ was determined from the average value of ${\bar
{\lambda_1}}$ obtained from the simulations.  This fitted value is in
excellent agreement with the analytic value of $2 / n \beta \Sigma =
0.02041\ldots$.  The form of (5.2) provides an excellent fit to the simulation
data with $\chi^2 = 0.8$.}
\oddsidemargin1.7cm
\evensidemargin1.7cm
\setlength{\textwidth}{18cm}
\end{figure}

In Fig.\ 4 we show a histogram of the distribution of the smallest eigenvalue
at $T= 0$ for the case $n=20$.  The results are
in perfect agreement with (4.1).
The temperature dependence of the parameters of the distribution
is shown in Fig. 5, where we show the ratio $\overline{\lambda_1}/\Delta
\lambda_1$ as a function of the temperature for $n = 10$.
When $x_0 = 0$ this ratio is
equal to $[\pi / (4-\pi) ]^{1/2} = 1.913\ldots$.  Surprisingly, we find that
that this ratio is constant for $T < T_c$.  In terms of $x_0$, this implies
that $x_0$ is strictly zero for $T < T_c$.  Inspection of the corresponding
distribution indicates that (5.1) remains quantitatively valid in this
region.  Above this temperature
(and ignoring some finite $n$ effects near threshold), $x_0$ grows linearly
with $T$.  This behavior is expected since the entire distribution
moves linearly with $T$ for sufficiently large $T$.  What is surprising is
that $x_0$ goes to zero below some finite $T$ and that the effect is so
pronounced for a matrix of such small dimension.

\vskip 0.5 cm
\noindent
\begin{figure}[h]
\begin{center}
\leavevmode
\psfig{file=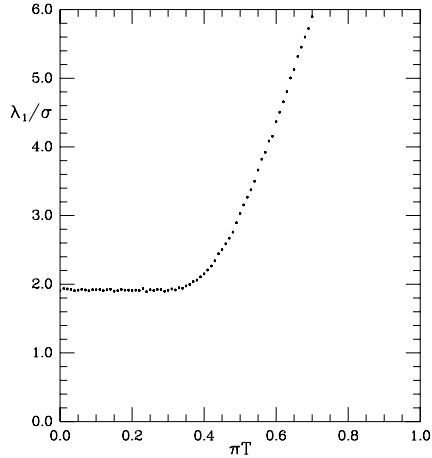,height=6cm}
\end{center}
\oddsidemargin3.7cm
\evensidemargin3.7cm
\setlength{\textwidth}{16cm}
\caption{ The ratio of the smallest eigenvalue to its rms deviation,
 $  \bar {\lambda_1}/\Delta \lambda_1$, as a function of temperature for
$n = 10$. Each point represents $ 2\times 10^4$ matrices. The average value
of this ratio for $T<T_c$ is 1.9107 which is in good agreement
with the expected value of
 $ (\pi/(4-\pi))^{1/2} = 1.913\ldots$. }
\oddsidemargin1.7cm
\evensidemargin1.7cm
\setlength{\textwidth}{18cm}
\end{figure}

The behavior of $x_0(T)$ versus $T$ (shown in Fig.\ 6) allows us to extract
$T_c$ from $x( T_c ) = 0$. In practice we use a linear extrapolation of
our data for $T > T_c$. Our results for different {\bf size} matrices can be
summarized by the expression
\be
\pi T_c^{(n)} = 0.40859 \ \frac{n + 2.51316}{n + 4.10346} \ \ ,
\ee
which yields an asymptotic result of $\pi T_c = 0.40859$, which is very close
to the theoretical result of $1/\sqrt{6} = 0.40824$.  As indicated, the form
of (5.2) is not exact.  Similar results can be obtained in a model-independent
fashion by straight-line extrapolation of the observed values of
$\overline{\lambda_1}/\Delta \lambda_1$ as a function of the temperature.
\vskip 0.5 cm
\noindent
\begin{figure}[h]
\begin{center}
\leavevmode
\psfig{file=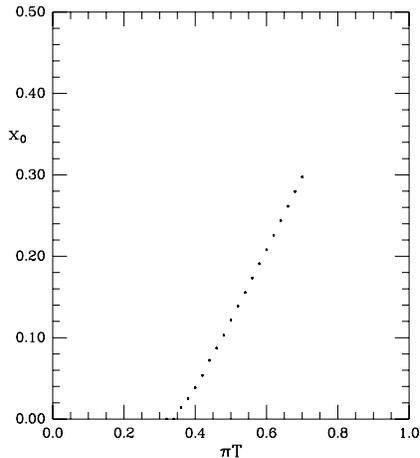,height=6cm}
\end{center}
\oddsidemargin3.7cm
\evensidemargin3.7cm
 \setlength{\textwidth}{16cm}
\caption{ The parameter $x_0$ of (5.2) as a function of $T$ for $n =10$.
Each point represents $2 \times 10^4$ matrices.  The parameters $x_0$ and
$\sigma$ were determined from the ratio ${\bar {\lambda_1}} /
\Delta \lambda_1$ (shown in Fig.\ 5) and the corresponding values of
${\bar {\lambda_1}}$.  Linear extrapolation suggests $\pi T_c = 0.3625$ for
this case. }
\oddsidemargin1.7cm
\evensidemargin1.7cm
\setlength{\textwidth}{18cm}
\end{figure}

Finally, we study the scaling of the smallest eigenvalue with $N$. For
$T< T_c$, the Banks-Casher formula suggests that $\lambda_{\rm min}\sim
1/n$. Indeed, this is what is found numerically both for the expectation
value and the variance of $\lambda_{\rm min}$.

At $T=T_c$ we have that $\Sigma(m) \sim m^{1/\delta}$. This leads to the
eigenvalue density (for $\lambda \rightarrow 0$)
\be
\rho(\lambda) \sim m^{\alpha_1} \lambda^{\alpha_2} \quad {\rm with} \quad
\alpha_1+\alpha_2 =\frac 1{\delta} .
\ee
Smaller masses  suppress the eigenvalue density near zero, so we must have
$\alpha_1 > 0$.  For a fixed mass, the eigenvalue density should
not diverge, so we must also have $\alpha_2 > 0$. It is possible
that $\alpha_1$ and $\alpha_2$ depend on $N_f$, but we were
not able to investigate this point within the present framework.
The scaling behavior of the smallest eigenvalue is obtained
from $\int_0^{\lambda_{\rm min}} \rho(\lambda) d\lambda \sim 1/N$ which leads
to
\be
\langle\lambda_{\rm min} \rangle\sim N^{-1/(\alpha_2 +1)}\ \ .
\ee
For $N_f=0$, the eigenvalue density cannot depend on $m$ and we thus have
\be
\alpha_2 = \frac 13\quad {\rm and }\quad \alpha_1 = 0 \quad{\rm for} \quad
N_f = 0 \ \ .
\ee
This yields $\langle\lambda_{\rm min} \rangle
\sim N^{-3/4}$, which has been verified
numerically to very high accuracy. It was also shown that the variance of
$\lambda_{\rm min}$ has the same scaling behavior.

For $ T>T_c$ the average position of the smallest eigenvalue departs from
zero and scales  like $N^0$.  We expect \cite{VZ} that its
variance well have the scaling behavior $N^{-2/3}$ which
is typical for eigenvalues near the edge of a semicircle. Again, this is
in perfect agreement with our numerical simulations.

\vskip 1.5 cm
\renewcommand{\theequation}{6.\arabic{equation}}
\setcounter{equation}{0}
\section{Conclusions}
In this paper we have studied a random matrix model possessing
the global symmetries of the QCD-action and the temperature
dependence suggested by the form of the lowest Matsubara frequency.
At $T=0$ this model is completely soluble; it reduces
to what is known in the mathematical literature as the Laguerre ensemble.
For nonzero temperatures, we have succeeded in extracting some
interesting properties analytically.  In particular, we have shown that
the model undergoes a second-order phase transition. Two critical exponents
were obtained.
\be
        \beta &=& \frac 12,\nonumber \\
        \delta &=& 3 \ \ .
\ee
The lattice result \cite{Karsch} for QCD with two light flavors
for $1/\beta\delta$ is $0.77\pm 0.14$,
which is closer to these mean field values than to the results for
either the $O(4)$ or $O(2)$ Heisenberg spin models. For three or
more light flavors, QCD shows a first-order chiral phase
transition, whereas for one flavor there is no transition at all. Our
model does not contain such flavor dependence: It has a
second-order phase transition for any
number of flavors.  If our model has anything to say about QCD, it
is for QCD with two light flavors.

Numerically, we found one surprising result: The
distribution of the smallest
eigenvalue below $T_c$ is (numerically) equivalent to the distribution
obtained for $T= 0$.  For $T>T_c$, the centroid of
the Gaussian distribution grows linearly with $T$.
The behavior at $T= 0$ agrees well with known analytical results.

Our results suggest an alternative method for obtaining the critical
temperature; namely from the distribution of the eigenvalue of smallest
magnitude. It would be interesting to study the dynamics of the smallest
eigenvalues in lattice QCD as well.

\vskip 1.5cm
\vglue 0.6cm
{\bf \noindent  Acknowledgements \hfil}
\vglue 0.4cm
 The reported work was partially supported by the US DOE grant
DE-FG-88ER40388. We would like to thank
H.A. Weidenm\"uller and T. Wettig for useful discussions and
for communicating their results on
a related model prior to publication.

\vfill
\eject
\newpage

\end{document}